**Possibility of obtaining atomic metallic hydrogen by electrochemical method**

**Nikolay E. Galushkin, Nataliya N. Yazvinskaya & Dmitriy N. Galushkin**

*Laboratory of electrochemical and hydrogen energy, Don State Technical University,*

*147 Shevchenko Street, Town of Shakhty, Rostov Region, Russia*

**Although hydrogen is the simplest atom, it forms extraordinary solids and liquids. It was shown in theoretical works, that hydrogen can transfer into metallic phase under high pressure[1]. Moreover resulting metallic hydrogen shall have metastable state at room temperature and pressure[2]. Metallic hydrogen shall possess a series of unique properties. First, its density shall be 10-13 times higher, than the density of liquid molecular hydrogen[3,4], i.e. it shall be the high capacity accumulator of hydrogen. Second, it shall be an ideal rocket fuel with specific energy 20 times higher than liquid hydrogen-oxygen fuel[3,4]. Third, at room temperature, it shall be a good conductor and possible that a superconductor[5]. However, it was not yet possible to obtain metallic hydrogen, in spite of the efforts of many researchers during more than 75 years. Here we demonstrate, that atomic metallic hydrogen (AMH) is formed inside of metal-ceramic oxide-nickel electrodes of nickel-cadmium battery over a long period of electrochemical hydrogenation (more than five years). We separated AMH from nickel metal by electrochemical dissolution of metal-ceramic nickel matrix, and researched its properties. It was established that density of AMH is 0.85 g cm$^{-3}$, released specific energy of hydrogen recombination is 216 MJ kg$^{-1}$, resistivity is 1.5 $\mu\Omega$ cm, and AMH is not a superconductor at room temperature. The obtained parameters of AMH coincide with the theoretical values forecasted earlier[3,4].**

In 1935, Wigner and Huntington predicted that pressures of order 25 GPa were required for the transition of solid molecular hydrogen to the atomic metallic phase[1]. Now considers that the dissociative transition will take place in the 400 to 600 GPa region[3,4]. This atomic metallic hydrogen (AMH) is predicted to be superconducting with a very high critical temperature, $T_c$,

>200 K (ref. 5). At lower pressures in the region 200-400 GPa, molecular hydrogen consisting of paired protons has been predicted to become a high temperature superconductor[6,7]. Theory predicts that metallic hydrogen might be a metastable material so that it remains metallic when pressure is released[2]. Calculations also predict that as solid molecular hydrogen is compressed, it transforms into a low-temperature quantum fluid before becoming a monatomic crystal[8,9].

Forecasted properties of AMH open up wide area of its practical application[3,4]. Density of AMH shall be within the limits of 0.7-0.9 g cm$^{-3}$, i.e. 10-13 higher, than the density of liquid molecular hydrogen. Thus, it might serve as a compact energy source for hydrogen economy (fuel cells, vehicles, etc) and as a lightweight building material. AMH recombines into molecular hydrogen with release of recombination energy of 216 MJ kg$^{-1}$. This is more than twenty times the specific energy released by the combustion of hydrogen and oxygen in the Space Shuttle's main engines, 10 MJ kg$^{-1}$ (ref. 4). Because of this very large potential specific energy, it might used as an energy source for rocket powered launch vehicles. If AMH is a superconductor with high critical temperature, then the area of its practical implementation in electrical engineering shall be very broad.

Hydrogen constitutes more than 90% of all atoms in the visible universe and contributes three quarters of its mass. It is widely accepted that hydrogen is abundant in the interiors of Saturn and Jupiter where it is both liquid and metallic, and the origin of their magnetospheres[10].

There are two methods of producing high pressures, static and dynamic. All modern studies on solid hydrogen at static high pressures have been carried out in diamond anvil cells[11-13]. In a dynamic shock-wave experiments achieve high pressures for very short periods of time, but the temperatures are thousands of degrees[14,15]. However in spite of the efforts of numerous researchers the metastable AMH is not yet obtained.

Therefore, a great deal of effort has been taken to seek an alternative route to hydrogen metallization. Experimental and theoretical works indicates that combination with tetravalent atoms, as in the group 14 hydrides[16-18], may significantly lower the metallization pressure of hydrogen. Hydrogen in these materials is "chemically precompressed" by the presence of the group 14 atoms in a crystal lattice. Our work expands this direction of metallic hydrogen research.

In our earlier study, it was demonstrated, that as a result of the thermal runaway of nickel–cadmium batteries, large amounts of hydrogen are released[19]. The thermal decomposition of electrodes demonstrated that hydrogen accumulates in the electrodes of nickel–cadmium batteries in the process of their operation. So KSX-25 battery with the service period of over five years contains approximately 800 liters of hydrogen. The capacity of an oxide-nickel electrode as a hydrogen absorber was quantified as 13.4 wt% and 0.4 g cm$^{-3}$ (ref. 20). The obtained result exceeds the earlier obtained results for nickel hydride (obtained using traditional methods) by 10 times[21], and for any reversible metal hydrides, including magnesium hydride or complex hydrides by 2 times[22,23]. This article is devoted to the determination of where and in what form hydrogen accumulates in electrodes of nickel–cadmium batteries.

It is possible to physically divide oxide-nickel electrode into two phases – active substance (nickel hydroxide) and metal–ceramic matrix (in the case of sintered electrodes present in KSX-25 batteries).

If hydrogen is intercalated into nickel hydroxides then, when nickel hydroxides reacts with acids and forms soluble salts, intercalary hydrogen will be released, because nickel hydroxides disappear and the salt dissolves into the solution. Any type of acid, which forms soluble salts with nickel hydroxide, but does not interact or poorly interact with metal matrix, can be used for this purpose. For example, it possible to use sulfuric acid, which interacts with nickel hydroxides with the formation of soluble salt of nickel sulfate.

Thus, if during the interaction of sulfuric acid with the oxide-nickel electrode, the hydrogen evolve in amounts, detected earlier in the experiments[19], then it shall confirm that hydrogen accumulates in nickel hydroxides during the operation of batteries. If the mass of metallic matrix after extraction of nickel hydroxides by sulfuric acid is higher than its mass before infiltration with hydroxides, it means that in the process of operation of the batteries, nickel hydroxides are transformed into some new stable chemical compounds, which are probably those that lead to the accumulation of hydrogen.

In this study, 22.6% sulfuric acid solution was used. In preliminary experiments, the rate of nickel hydroxide dissolution in oxide-nickel electrode over sulfuric acid was installed. The experiment showed that after 15-20 minutes of etching, the electrode mass practically does not change. For the purpose of reliability, the period of oxide-nickel electrodes etching was taken as 30 minutes. The experiment also confirms a well-known fact that sulfuric acid very weakly interacts with metallic nickel.

To verify the hypothesis of hydrogen accumulation in nickel hydroxide, 10 oxide-nickel electrodes from different KSX-25 batteries were selected. After washing in distilled water and drying, they were weighed and placed in acid in the flask for 30 minutes. The flask was tightly closed by a rubber plug with a tube for removal of gas. Gas was collected into the gas receiver. The gas receiver allowed monitoring evolution of gas, with the accuracy of up to 10 ml (ref. 19). Then, electrodes were dried and weighed again. Loss of mass by oxide-nickel electrodes as a result of their etching in sulfuric acid was 32 – 36%. According to manufacturers' information, the positive electrode contains 30–34% of nickel hydroxides and 1.5–2% of cobalt hydroxides. The results obtained for the mass loss of electrodes are found to be exactly within this range. Consequently, nickel hydroxides in the process of batteries' operations are not transformed into other stable chemical compounds that are capable of hydrogen accumulation. Hydrogen does not evolve at all within the sensitivity limits of installation (10 ml). It unequivocally follows, that no hydrogen intercalated into nickel hydroxide.

Thus, there is only one possibility left – hydrogen accumulates in metal-ceramic oxide-nickel matrix during the operation of the KSX-25 batteries. Transition elements, to which nickel refers, are capable of accumulating hydrogen[21-23]. Theoretically, three situations are possible: hydrogen is intercalated into metal-ceramic nickel matrix; hydrogen locates in micro-defects of metal-ceramic nickel matrix; hydrogen forms chemical compounds with metal-ceramic nickel matrix. Therefore, if to assume, that hydrogen is intercalated into the metal-ceramic matrix or locates in its micro-defects, hydrogen shall evolve from the electrochemical dissolution of nickel matrix. However, if to assume, that hydrogen forms chemical compounds, then sediments get deposited during the electrochemical dissolution of electrode under the condition that the compounds itself do not disintegrate chemically or electrochemically in the given solution.

Electrochemical dissolution of nickel matrix was performed in a sealed flask described above. The solution of the following composition was used in experiments: $NiSO_4 * 7H_2O$ – 200 g $l^{-1}$; $Na_2SO_4$ - 80 g $l^{-1}$; $NaCl$ – 40 g $l^{-1}$; $H_3BO_3$ – 15 g $l^{-1}$. Nickel matrix dissolution was executed at the current density 2.5 A $dm^{-2}$.

The results of experimental research were not as expected. As a result of the metal-ceramic nickel matrix dissolution, the portion of electrode submerged into solution, detached, and dropped to the flask bottom. The fallen portion had the appearance of the initial electrode without any changes. However during extraction from the solution and when the fallen portion was touched, it split into small crystals of grey color with metallic luster. It is possible to suppose that the dropped portion of investigated electrode represents the crystals' agglomeration of nickel hydride. As a result of electrochemical dissolution of metal-ceramic matrix of the oxide-nickel electrode, evolution of hydrogen was not detected, within experimental error (10 ml). The given result was not as expected; as observed in any type of hydrogen accumulation in metal, the first stage is the intercalation of hydrogen into metal. It is possible to explain the absence of hydrogen intercalated into metallic matrix by degasation of oxide-nickel electrode during storage of batteries after their battery life is over. All the batteries investigated in this experiment were stored after their battery life is over for not less than one year. Experiment show that the oxide-nickel electrodes of nickel-cadmium batteries KSX-25 with a long life consists of three phases: active substance (nickel hydroxide), chemical compound, of supposedly, nickel hydride, and pure nickel, approximately in equal mass fractions. Hence, the experiment shows that hydrogen is accumulated in powder (supposedly in nickel hydride) a bound state.

Firstly, we evaluate the specific parameters of the powder as a hydrogen absorber. An oxide-nickel electrode accumulates approximately 36 liters of hydrogen during the long service life (over five years) of a KSX-25 battery[19,20]. Weight of the oxide-nickel electrode is 24 g, and the weights of nickel hydroxide, nickel hydride powder, and pure nickel equal to 8 g each. Hence, the specific mass capacity of an oxide-nickel electrode as a hydrogen absorber equals to 13.4 wt% (ref. 20), metal-ceramic nickel matrix 20.1 wt% and powder 40.2 wt%. Considering that the physical dimensions of an oxide-nickel electrode of a KSX-25 battery equal to $7.3 \times 13.6 \times 0.081$ cm, we obtain that the specific volume capacity of an oxide-nickel electrode as a hydrogen absorber equals to 0.4 g cm$^{-3}$ (ref. 20) and powder 0.85 g cm$^{-3}$; because the volume of powder is 2.13 times smaller than the volume of electrode.

From the specific mass capacity for powder we obtain the formula $NiH_{39.4}$. Thus the powder is composed by 97.5% of hydrogen atoms and 2.5% of nickel atoms. Hence the powder is not nickel hydride, but dense hydrogen with small impurity of nickel atoms. The powder is a good conductor. Its resistivity is 1.5 µΩ cm. Moreover resistance decreases with growth of pressure, which is related to better contact between the grains of powder. Hence, the powder is a metastable phase of metallic hydrogen. Unfortunately the powder is not a superconductor at room temperature ($T$=298 K).

In our earlier study[19], on the base of analysis of an energetic balance of a thermal runaway for a KSX-25 battery, it was shown that in the process of the thermal runaway, the battery emits energy at least 6 times that much than it obtains from the recharger. Moreover, in this work was estimated only the lower limit of the energy emitted. In fact, the energy released during the thermal runaway is much more. Thus, the results of the estimation of the thermal runaway energy balance unambiguously show that thermal runaway is accompanied by a powerful exothermic reaction with the release of a great amount of heat.

In this work we performed calorimetric analysis of thermal runaway (similar to the one, which was under research in the paper ref. 19). Experimental research showed, that in the result of thermal runaway the energy equal to 5232 kJ (relative error 5%) released from the accumulator. Meanwhile the accumulator receives from charging device the energy equal to 33.7 kJ (ref. 19) (during the process thermal runaway). Besides 270 liters of hydrogen evolves in the process of thermal runaway. Usually during such a thermal runaway all electrolyte evaporates and the plastic body of the battery melts down.

Supposing that hydrogen in the electrode is in atomic phase, in case of its recombination and evolving of 270 liters of hydrogen, the energy of 5207 kJ shall release. This value well corresponds to the above experimental value obtained. Hence, hydrogen is present both in the oxide-nickel electrode, and the above obtained powder in atomic phase. Hence, the powder is the metastable phase of atomic metallic hydrogen (AMH).

Hydrogen evolves from the battery in the process of thermal runaway during 2-4 minutes. Hence, the temperature of electrodes in the place of thermal runaway is a critical temperature of the metastable phase of AMH transfer into molecular gas. Thermal runaway takes place not in all the surface of electrodes, but in local areas. During this process the separator burns through in the place of thermal runaway in the form of the circles of different radius (depending on the power of thermal runaway). Thermal runaway develops at electrode surface at random locations. That is why it is very difficult to determine the exact temperature of electrodes in the place of thermal runaway. Our measurements show, that $T_c$>1200 K.

Let us consider the causes of accumulation of such a big quantity of hydrogen in oxide-nickel electrodes of alkaline batteries. It must be noted that in an oxide-nickel electrode there are a number of factors present, which contribute to hydrogen accumulation. At present, these factors are under intensive research.

First, the nickel oxide electrode in a KSX-25 battery is metal-ceramic, made from finely divided nickel powder with strongly collapsed crystalline structure. Any imperfections of metal crystalline structure (particularly dislocations) are traps for hydrogen, as they decrease the energy of hydrogen atom as compared to their location in normal interstice. Besides they are the centers of hydrogen absorption, and also contribute to hydrogen penetration into the metal depth. Hence, imperfections of the metal crystalline structure cause sharp rise of hydrogen miscibility in it. The hydrides used in the modern methods of preparation are ground down in ball mills with the above mentioned purpose[24,25].

Secondly, an oxide-nickel electrode contains nickel oxides. It is a well known fact that the oxides of transition metals act as catalysts of hydrogen accumulation[26,27].

Thirdly, the electrodes are densely packed. Thus, hydrogen evolved at the cadmium electrode during charging of the battery may penetrate into the pores of both the oxide-nickel, and cadmium electrodes. Hence, in all probability, the oxide-nickel electrode accumulates hydrogen not electro-chemically, but due to the high capillary pressure. It was shown in paper[19], that capillary pressure in KSX-25 battery electrodes can reach 100 MPa and more.

Lastly, the process of hydrogen accumulation in the batteries' electrodes takes place all through their service life. This is more than six years in the present study. The KSX-25 battery operates in the mode of floating charge. Hence, it overcharges during most of the time. During this process electrolyte decomposes into hydrogen and oxygen. Hydrogen is characterised by a very high diffusion permeability. For example, at the temperature 20 $^0$C the diffusivity of hydrogen in nickel is approximately $10^{10}$ times higher than diffusivity of nitrogen or oxygen[23]. These features explain why upon dissociation of an electrolyte into hydrogen and oxygen in a battery, only hydrogen accumulates in the electrodes, while oxygen escapes to the atmosphere.

With the modern methods, the process of hydrides preparation could take from several minutes to several hours. Analysis of hydrides' research works for approximately last 20 years showed, that most probably there were no experiments for obtaining transition metals hydrides at high external hydrogen pressures lasting for extremely long time periods. However exposure to external pressure is a significant factor at any processes of penetration of one substance into another. Especially when the penetration process is extremely slow as in our case.

Duration of electrodes' hydrogenation process is of very high importance for formation of AMH. It was shown in paper[1], that for formation of AMH the first stage is very important, i.e. the formation of nuclei AMH. Potential energy barrier separating molecular hydrogen from AMH is very high. It is exactly due to this fact, that AMH was not yet obtained. However, inside the metallic matrix of electrode there is quite a number of factors decreasing the above barrier and contributing to formation of nuclei AMH. First, upon accumulation of quite big amount of hydrogen in metal-ceramic nickel matrix, it starts to form chemical compounds with nickel (β-phase of nickel hydride). Hydrogen in this phase is already in atomic state, just like in AMH[23]. Secondly, hydrogen in these materials is "chemically precompressed" by the presence of the atoms (Ni) in a crystal lattice and by the presence of new intercalating hydrogen atoms[16-18]. Thirdly, second component (Ni), seen as an impurity, may potentially reduce the pressure required to metallization hydrogen[28]. Since the nickel atoms can be centers of formation of nuclei AMH (as in AMH powder obtained by us). AC and DC fields in the battery also contribute to the formation of nuclei AMH phase[29]. In spite of the above favorable factors the probability of nucleation of a new AMH phase in an oxide-nickel electrode is extremely small. That is why long time is required for formation of big quantity of nuclei AMH phase (as in the obtained AMH powder). Besides the average radius of nuclei grows with time $t$ according to the law $r = At^{1/3}$, where $A$ is an extremely small value, determined by diffusion of bound hydrogen atoms in β-phase nickel hydride[30]. Hence, for formation relatively big nuclei AMH long process hydrogenation is required.

We verified experimentally the above theoretical conclusions. It was shown in paper[20], that hydrogen is absent in the electrodes of new nickel-cadmium batteries, but during its service life the amount of hydrogen inside the electrodes increases. The amount of hydrogen absorbed by the electrodes stops increasing after five years of service life, i.e., the maximum capacity of the electrodes for hydrogen storage is reached after that time.

In this work we researched the amount of AMH powder in oxide-nickel electrodes of nickel-cadmium batteries with different service life. It was established that there was no powder in the electrodes of new batteries. As service life of the batteries grows, the amount of AMH powder in oxide-nickel electrodes grows. However, in the electrodes of batteries with service life of up to one year, there was practically no AMH powder. Considerable amount of AMH powder appears only after three years of battery operation; and its maximal amount is achieved after five years of service life. The above experimental research explicitly prove, that duration of hydrogenation process of oxide-nickel electrodes is a decisive factor for AMH phase formation. The AMH problem, of course, requires further research, both experimental and theoretical.

---

**Competing interests statement**. The authors declare that they have no competing financial interests.

**Correspondence** and requests for materials should be addressed to N.E.G. (galushkinne@mail.ru).